\documentclass[aps,prl,twocolumn,superscriptaddress]{revtex4-1}
\usepackage{graphicx,color}
\graphicspath{ {./figuras/} } 

\begin{document}

\title{Influence of relativistic effects on the contact formation of transition metals}

\author{M. R. Calvo}
\email{Electronic mail address: r.calvo@nanogune.eu}
\affiliation{Departamento de F\'\i sica Aplicada and Unidad asociada CSIC, Universidad de Alicante, Campus de San Vicente del Raspeig, E-03690 Alicante, Spain.}
\affiliation{Ikerbasque, Basque Foundation for Science, 48013 Bilbao, Spain}
\affiliation{CIC nanoGUNE, 20018 Donostia-San Sebastian, Spain}

\author{C. Sabater}
\altaffiliation[Present address: ]{Chemical Physics Department,
Weizmann Institute of Science,
 76100 Rehovot, Israel.}
\affiliation{Departamento de F\'\i sica Aplicada and Unidad asociada CSIC, Universidad de Alicante, Campus de San Vicente del Raspeig, E-03690 Alicante, Spain.}

\author{W. Dednam}
\affiliation{Departamento de F\'\i sica Aplicada and Unidad asociada CSIC, Universidad de Alicante, Campus de San Vicente del Raspeig, E-03690 Alicante, Spain.}
\affiliation{Department of Physics, Science Campus, University of South Africa, Florida Park, 1710, South Africa}

\author{E. B. Lombardi}
\affiliation{College of Graduate Studies, University of South Africa, Pretoria, 0003, South Africa}

\author{M. J. Caturla}
\affiliation{Departamento de F\'\i sica Aplicada and Unidad asociada CSIC, Universidad de Alicante, Campus de San Vicente del Raspeig, E-03690 Alicante, Spain.}

\author{C. Untiedt}
\email{Electronic mail address: untiedt@ua.es}
\affiliation{Departamento de F\'\i sica Aplicada and Unidad asociada CSIC, Universidad de Alicante, Campus de San Vicente del Raspeig, E-03690 Alicante, Spain.}

\date{\today}

\begin{abstract}

Our analysis of the contact formation processes undergone by Au, Ag and Cu nanojunctions, 
reveals that the distance at which the two closest atoms on a pair of opposing electrodes 
jump into contact is, on average, two times longer for Au than either Ag or Cu. This suggests the existence of a longer range interaction between those two atoms in the case of Au, a result of the significant relativistic energy contributions to the electronic structure of this metal, as confirmed by \textit{ab initio} calculations. Once in the contact regime, the differences between Au, Ag and Cu are subtle, and the conductance of single-atom contacts for metals of similar chemical valence is mostly determined by geometry and coordination.

\end{abstract}


\maketitle
As the atomic mass increases, relativistic effects come into play that modify the electronic structure and thus determine the properties of heavy-element crystals and compounds (for a review see e.g. Ref.  \cite{pyykko1988relativistic}). As a consequence of this, $5d$ metals differ markedly from their $4d$ counterparts; examples include Hg being liquid at standard conditions \cite{Calvo_2013} and the golden luster of bulk Au \cite{pyykko1988relativistic}.
In $5d$ transition metals, relativistic energy contributions result in a contraction of the outer $6s$ shell, accompanied by the expansion of the filled $5d$ orbitals \cite{pyykko1988relativistic}, producing an enhancement in the $s$-$d$ hybridization of the valence orbitals. This has a strong influence on the bonding properties of $5d$ atoms and thus determines the chemistry of the $5d$ elements \cite{pyykko1988relativistic,Gorin_2007}, as well as many physical properties of their bulk crystals. 
Au exhibits distinct mechanical and structural properties such as a larger bulk moduli and cohesive energies than Ag \cite{ho1987stability}. This effect is further amplified in low-coordination structures, where it gives rise to phenomena such as 
surface reconstruction\cite{takeuchi1989theoretical,filippetti1997reconstructions}. A similar origin is attributed to the formation of monoatomic chains during the rupture of  Au, Pt and Ir nanostructures \cite{Smit_2001}, in agreement with expected enhanced s-d hybridation of the valence orbitals in these one-dimensional structures \cite{hakkinen2000nanowire,Bahn_2001,Hasmy_2008,thiess2008theory}. 

 The process of contact formation at the nanoscopic scale has been the object of extensive study in the context of single-atom metallic junctions \cite{agrait2003quantum,Kr_ger_2009}. 
 The influence of geometry and coordination on both the contact formation process as well as the characteristics of the resulting structures, are well-established for a variety metals \cite{Untiedt_2007,sabater2013understanding,Limot_2005,Trouwborst_2008}.
 In particular, for Au, Ag and Cu, it is well known that first contact in low-coordination geometries is invariably accompanied by an abrupt jump
\cite{Untiedt_2007,Fern_ndez_2016}, upon which a single or double atomic contact is normally found to have formed immediately afterward. 

In this work, the formation of thousands of atomic contacts made of pure Au, Ag and Cu, is studied and compared. 
 In the contact regime, and in agreement with previous works, the conductance is mainly determined by the valence of the metal \cite{scheer1998signature} and the exact geometry of the contacts \cite{agrait2003quantum,Untiedt_2007,Trouwborst_2008,sabater2013understanding}.
 In the tunneling regime, on the contrary, we find that there is a remarkable difference between Au, on one hand, and Ag and Cu, on the other, in terms of the distance from which jump to contact starts, which we show here to be a consequence of the larger relativistic effects in the electronic structure of Au. 


\begin{figure}[htp]
\includegraphics[width=0.45\textwidth]{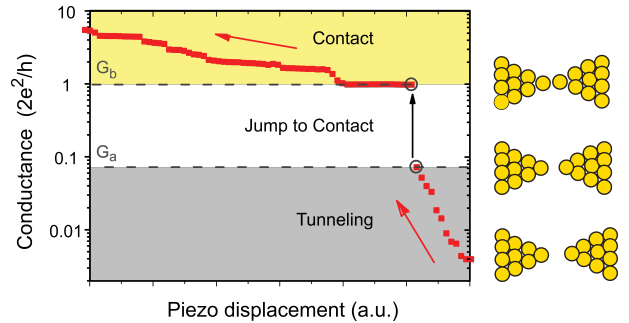}
\caption{A trace of conductance recorded during the formation of a gold contact in a STM setup at 4.2K. As the electrodes approach each other, an initial exponential increase in conductance is followed by a jump into contact, as indicated by the vertical black arrow. Conductance values before and after contact formation are marked and labeled as G$_{a}$ and $G_{b}$ respectively. On the right, from bottom to top, different stages of the contact process are illustrated.}
\label{Fig1}
\end{figure}

Our atomic contacts are fabricated by cyclic loading of two electrode probes made of the same high purity (99.999\%) metal,  Au, Ag or Cu, under cryogenic vacuum at 4.2K.  The electrical conductance (obtained as the current divided by the applied voltage of 100 mV) is recorded while the two electrodes are carefully brought into contact in a Scanning Tunneling Microscope (STM) setup. The process is described in detail in previous works \cite{Untiedt_2007,sabater2013understanding}. Traces of conductance, such as the one shown in Fig.~\ref{Fig1}, can be obtained in this way. When the atomic-sized electrodes are close enough but not yet in contact, electrons may tunnel from one to the other. In this tunneling regime,  the conductance increases exponentially with decreasing distance between the leads. This increase in conductance remains smooth until a sudden jump occurs, and a plateau at around the value of one quantum of conductance $G_{0}=2e^2/h$ appears, indicating the formation of a monoatomic contact \cite{agrait2003quantum}. 

\begin{figure}[htp]
\centering
\includegraphics[width=0.5\textwidth]{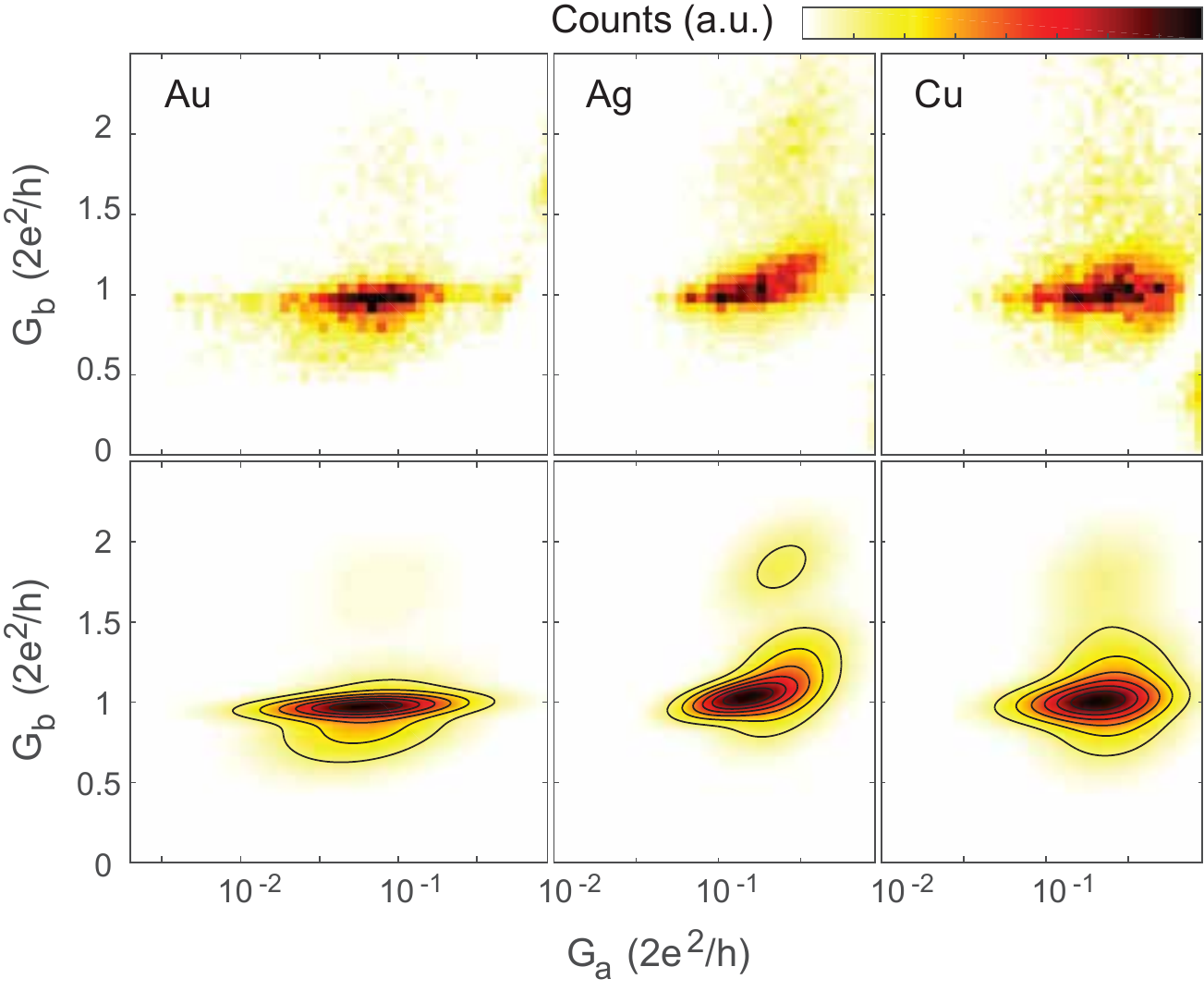}
\caption{Upper panels: Density plots of jump to contact parameters extracted from more than 2000 contact formation traces for Au, Ag and Cu (labeled correspondingly). Bottom panels: evaluation of the results from fitting the data in the upper panels to a sum of three bivariate normal distributions.}
\label{Fig2}
\end{figure}

For each contact-formation trace, we search for the largest jump in conductance between two consecutive points. Thus, two conductance values are recorded, $G_{a}$, from which the jump occurs, and $G_{b}$, the final value immediately after the jump, labeled accordingly in Fig.~\ref{Fig1}. Following the analysis introduced by Untiedt et al. \cite{Untiedt_2007}, we construct density plots from this set of data pairs $(G_{a},G_{b})$. The top row of panels in Fig.~\ref{Fig2} represents such density plots, compiled from more than 2000 contacts formed by Au, Ag and Cu, respectively.
Density maxima appear at the most probable values of $(G_{a},G_{b})$ from and to which the conductance jump occurs.

The results for Au, presented in the left panels of Fig. 2, are similar to those reported in previous works \cite{Untiedt_2007,Trouwborst_2008,sabater2013understanding}, except that here we have plotted the $G_a$ data on a logarithmic scale. This, which could be seen as just a subtle change, enables a new and improved analysis of these data. First, the log-scale conveys a more physical interpretation of our results, since $log(G_a)$ \footnote{log denotes here the common logarithm (base 10)} is directly proportional to the distance between electrodes. Note that the conductance in the tunneling regime depends exponentially on the distance between the electrodes as $G\simeq Ke^{-\frac{\sqrt{2m\phi}}{h}d}$, where $K$ is a proportionality constant which depends on the area and density of states at the Fermi level of the electrodes, $m$ corresponds to the electron mass and $\phi$ is the metal work function. Therefore, a change in conductance of one order of magnitude corresponds roughly to a variation in distance of 1 \AA.

Moreover, Fig.2 reveals relevant information on the statistical distribution of data, unaccessible before. When plotted on a $G_a$ linear scale as in Ref. \cite{Untiedt_2007} density plots exhibit a triangular shape, allowing only for a rough identification of distributions and their most probable values on $G_a$. In contrast, on a logarithmic scale (Fig.2), 
density plots resemble normal distributions in both $G_b$ and $log(G_a)$. In fact, the density maximum around the quantum of conductance can be modeled as the superposition of two distributions. A third maximum, associated with a lower number of counts, can also be observed at higher values of $G_b$.  Hence, our data can be fitted to the sum of three bivariate normal distributions, (see \cite{PRB} for details), and thus allowing for a more precise identification of distributions as well as their mean and standard deviation values. Fit results are shown in the lower panels of Fig. 2 and graphically summarized in Fig. 3. In a related work (see Ref. \cite{PRB}), we show how these three maxima correspond to three first-contact configuration categories, namely, monomers, dimers and double contacts, in agreement to previous works \cite{Untiedt_2007,sabater2013understanding,Fern_ndez_2016}. However, in contrast to those works, the distributions corresponding to monomers and dimers can be now clearly disentangled. Furthermore, in Ref. \cite{PRB} we also present a more precise identification of contact geometry and conductance based on our analysis. 

\begin{figure}[htp]
\includegraphics[width=0.5\textwidth]{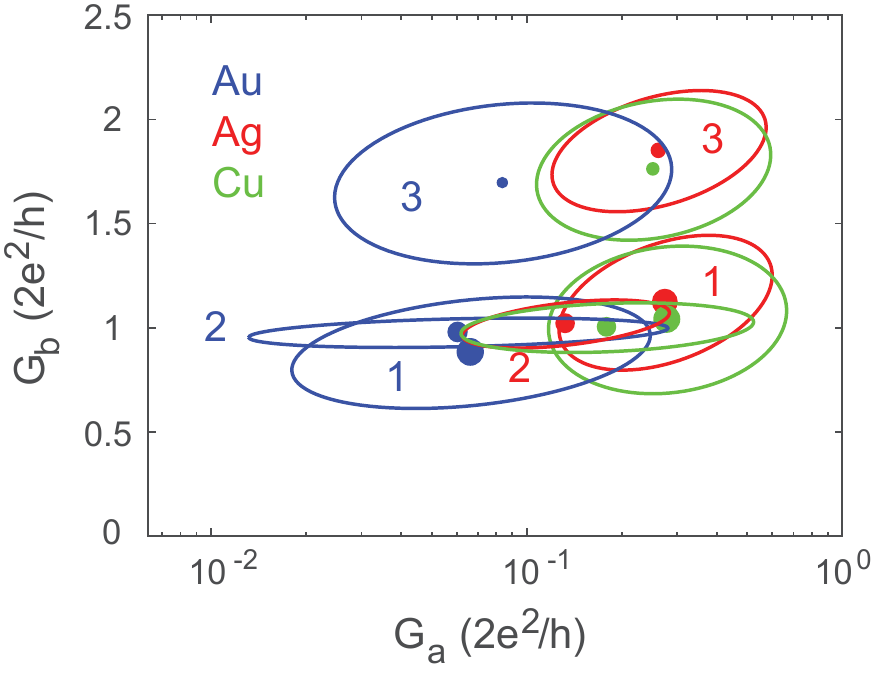}
\caption{Graphical summary of the fitting results used to compare the metals. Dots mark the mean value of the distribution, the size of which is proportional to the ratio of occurrence of each configuration, and the ellipse contour encloses the equivalent of one standard deviation for each bivariate distribution, in other words, the ellipse encloses  68\% of the data.}
\label{JClogAuagCu}
\end{figure}


A comparison of the results for Au, Ag and Cu contacts reveals a striking difference between Au and the other two metals, in the distribution of $G_a$ values. As listed in Table I and clearly visible in Fig. 3, the mean value of the distance over which Au jumps into contact (proportional to $log(G_a)$) is much larger in magnitude, for all three of its associated distributions. Assuming that in tunneling an increase of an order of magnitude in conductance corresponds approximately to a change of 1 \AA \space in distance, jump to contact for gold occurs at distances up to $\sim 0.5$ \AA \space larger than Ag and Cu, with a broader distribution. 
This compares favorably with the mean \textit{binding lengths} calculated for Au and Ag junctions from experimental force-extension curves in Ref. \cite{Hybertsen_2016}, which in the case of Au is $\sim 0.8$ \AA \space longer.
At the same time, the conductance at first contact exhibited by Au is slightly smaller than for Ag and Cu, which, in turn, exhibit similar values. 

\begin{table}
\begin{tabular}{|p{0.7cm}||p{2cm}|p{2cm}|p{2cm}|}
 \hline
 \multicolumn{4}{|c|}{\bf{$\mu_{log(G_a/G_0)} (\pm\sigma_{log(G_a/G_0)})$}} \\
 \hline
 &{\textcolor{blue}{\textbf{Au}}} & {\textcolor{red}{\textbf{Ag}}}& {\textcolor{green}{\textbf{Cu}}} \\
 \hline
 \bf{1} & -1.2 $\pm$ 0.4 & -0.6 $\pm$ 0.2 & -0.6 $\pm$ 0.2\\
 \bf{2} & -1.2 $\pm$ 0.4 & -0.9 $\pm$ 0.2 & -0.8 $\pm$ 0.3 \\
 \bf{3} & -1.1 $\pm$ 0.4 & -0.6 $\pm$ 0.2 & -0.6 $\pm$ 0.2 \\
  \hline
\end{tabular}
\caption{Mean values of log($G_{a}/G_{0}$) ($\pm$ standard deviation) values extracted from the fitting of Au, Ag and Cu density plots to three bivariate distributions, labeled as 1,2 and 3 in Fig.3 and associated respectively 
to monomer, dimer and double bond geometries \cite{Untiedt_2007,PRB}. $G_0$=$2e^2/h$ denotes here a quantum of conductance.}
\end{table}

All of the above can be understood in terms of a longer range interatomic potential felt by the atoms on opposing Au electrodes, as compared to electrodes made of Ag or Cu. In the case of gold, this interaction manifests much sooner, as the force required relative to the bulk elasticity to provoke the jump to contact. The stronger interaction also implies a smoother variation of the interatomic potential as a function of the separation between the Au electrodes, which explains the broader distribution in $G_a$ values that is observed. Finally, the fact that Au jumps to contact earlier, produces strained structures exhibiting a somewhat lower conductance, which in the case of a dimeric configuration gives way to the narrower distribution of $G_b$ values seen in Fig. 3. 

Hence, the longer range interaction would then seem to explain all the observed features of gold. Since Au, Ag and Cu share very similar electronic configurations, one can expect the long range interaction here to originate from relativistic effects, as these are responsible for other similar physical properties in which Au differs from Ag and Cu \cite{ho1987stability,takeuchi1989theoretical,filippetti1997reconstructions,Smit_2001}, as previously explained.

To test the above hypothesis, we have performed \textit{scalar} relativistic and non-relativistic total-energy density functional theory (DFT) calculations on infinite monatomic chains of gold and silver \cite{Bahn_2001}. For this, we have employed the plane-wave DFT code CASTEP \cite{clark2005first}, explicitly including or excluding scalar relativistic interactions. We make use OTFG pseudopotentials \cite{Vanderbilt_1990}  
(benchmarked against fully converged all-electron DFT calculations, with an error of 0.5 meV/atom obtained by the methods described in Ref. \cite{Lejaeghere_2013}). The (scalar) relativistic treatment is at the level of the Koelling-Harmon approximation of the Dirac equation \cite{Koelling_1977}, which, with the exception of spin-orbit coupling (SOC), retains all other relativistic kinematic effects such as mass-velocity, Darwin, and higher order terms. Since monatomic chains and atomic point contacts made of gold do not appear to exhibit significant local magnetic order \cite{Strigl_2015}, we have neglected SOC in our calculations. As exchange-correlation functional, we have used the generalized gradient approximation (GGA) by Perdew-Burke-Ernzerhof (PBE) \cite{perdew1996generalized}. We used the Tkatchenko-Scheffler (TS) dispersion-correction scheme to take van der Waals interactions between atoms into account. We have also used the plane-wave cut-off in Ref. \cite{Hybertsen_2017} for gold, 400 eV, whilst silver required a larger value, 600 eV. Convergence was checked with respect to plane wave
cutoff, with total energies converged to within $5 \times 10^{-7}$ eV/atom.
A total of 24 irreducible k points were used to sample reciprocal space in our calculations. Convergence was also checked for k points by gradually increasing the size of the Monckhorst-Pack grid automatically generated by CASTEP. To speed up our calculations, the symmetry was restricted to P4/mmm.

\begin{figure}[htp]
\includegraphics[width=0.49\textwidth]{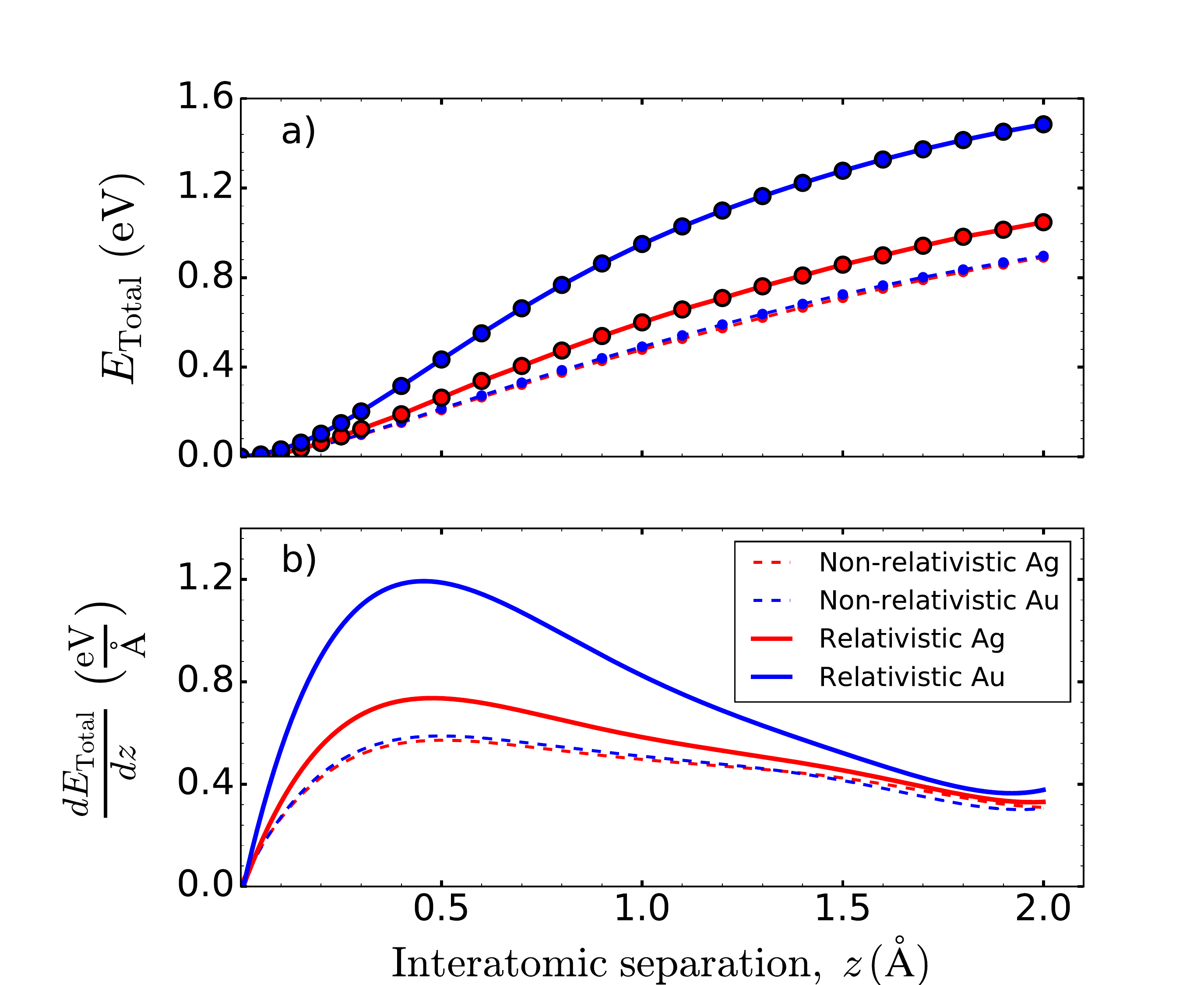}
\caption{(Rescaled) total energy a), and its derivative b), as a function of (rescaled) interatomic separation, $z$,  between atoms in infinite monatomic chains of non-relativistic gold (dashed blue) and silver (dashed red), and scalar relativistic gold (solid blue) and silver (solid red). The derivative curves in b) were obtained from a fit of the total-energy data in a) to sixth-order polynomials. In the absence of scalar relativistic corrrections, monatomic gold and silver chains exhibit almost identical force curves. Conversely, the interaction ``force" between relativistic gold atoms in b) rises to the maximum value for relativistic silver atoms a distance $\sim 0.7$ \AA \space earlier.}
\label{EnergyRelaNrelAuAg}
\end{figure}

Each unit cell of the infinite chain contained one atom, with the chain oriented along the $z$-axis ($c\sim 2.5$ \AA), and, to avoid interactions between periodic images, at least 10 \AA \space vacuum in the $\hat{x}$ and $\hat{y}$ directions ($a=b=10$ \AA) \space \cite{Hybertsen_2017}. We first optimized the interatomic separation between the individual atoms in the chain by varying the cell size along $z$, while keeping all other dimensions fixed, until the per-atom force fell below $10^{-2}$ eV/\AA. We used the TPSD algorithm \cite{barzilai1988two} for constrained relaxations. Then, starting from the equilibrium separation, we increased the interatomic spacing within the chains, incrementing $c$ by 0.1 \AA \space at a time, and calculated the total energy as a function of interatomic separation, similar to Ref. \cite{Hybertsen_2017}. Figure \ref{EnergyRelaNrelAuAg} a) shows the results of these calculations.

The total energy in Fig. \ref{EnergyRelaNrelAuAg} a) clearly rises more steeply with interatomic separation in the case of relativistic gold (in the figure, the origin coincides with the equilibrium interatomic separations and corresponding energies of the chains). To obtain an estimate of the interaction ``force" between the atoms in the chains, as a function of separation between them, we have fitted the total-energy data in Fig. \ref{EnergyRelaNrelAuAg} a), to sixth-order polynomials, and then took the derivative of the results (see Fig. \ref{EnergyRelaNrelAuAg} b) ). The maxima of the derivative curves in Fig. \ref{EnergyRelaNrelAuAg} b) can be identified with the ``force" required to break the monatomic chains \cite{Bahn_2001,Hybertsen_2017}. The obtained values, $\sim 1.9$ and $1.2$ nN for relativistic Au and Ag, respectively, exhibit good agreement with experiment and previous calculations \cite{Trouwborst_2008,Hybertsen_2016}. 

It is remarkable that the relativistic gold atoms already experience an interaction ``force", equal in magnitude to the maximum ``force" between relativistic silver atoms, when the separation between them is $\sim 0.7$ \AA \space greater. In particular, in the absence of scalar relativistic corrections, the force curves are virtually identical. In reality, there is a small horizontal offset ($\sim 0.1$ \AA, not shown) between these two force curves, due to their slightly different equilibrium chain lengths. Furthermore, based on elastic constants alone, one would expect silver to jump earlier than gold, since the difference in Young's moduli of these metals in the bulk is enhanced in low-coordination environments (e.g, in exposed surface layers \cite{Hofer_2004}, or as adatoms on free surfaces \cite{Limot_2005}). Silver is thus associated with a lower ``elastic constant" and is, therefore, ``softer". Of course, the size of the jumps cannot be entirely explained by these ``intra-electrode" elasticity arguments. Rose's universal binding potential \cite{rose1981universal}, $U=-\alpha(x-x_0)e^{-\beta(x-x_0)}+E_{0}$, a simple model of competing inter- and intra-electrode atomic interactions, suggests that a stronger inter- versus intra-electrode binding is responsible for the jump to contact phenomenon \cite{Trouwborst_2008}. This model relates the equilibrium bond length $x_{eq}=x_{0}+1/\beta$ and energy $E_{\mathrm{eq}}=E_{0}-\alpha/e\beta$ to a number of fitting parameters of physical importance, such as the breaking force $F_{\mathrm{break}}=-\alpha/e^2$ and inter-electrode binding constant $k_{\mathrm{bind}}=-\alpha\beta/e^3$. Table II records the fits of our DFT total energy data to this model.

\begin{table}
\begin{tabular}{|p{2.2cm}||p{1.4cm}|p{1.4cm}||p{1.4cm}|p{1.4cm}|}
 \hline
 & Au no rel. & {\textcolor{blue}{\textbf{Au rel.}}}&  Ag no rel. & {\textcolor{red}{\textbf{Ag rel.}}}\\
 \hline
 $\bf{\alpha}$ (eV/\AA) & 4.48 & 8.66 & 4.34 & 5.46\\
 \hline
 $\bf{\beta}$ (1/\AA) & 1.58 & 1.99 & 1.45 & 1.66\\
 \hline
 $\bf{x_{0}}$ (\AA)& 2.2 & 2.1 & 2.05 & 2.052 \\
 \hline
 $\bf{E_{0}}$ (eV) & -14080 & -14130 & -3970 & -4004 \\
 \hline
 \hline
 $\bf{x_{eq}}$ (\AA) & 2.86  & 2.58 & 2.74 & 2.65\\
 \hline
 $\bf{E_{eq}}$ (eV) & -1.084  & -1.60 & -1.10 & -1.21 \\
 \hline
 \hline
 $\bf{x_{break}}$ (\AA) & 3.52 & 3.09 & 3.43 & 3.26 \\
 \hline
 $\bf{F_{break}}$ (eV/\AA) & -0.61 & -1.17 & -0.59 & -0.74 \\
 \hline
 \hline
 $\bf{x_{bind}}$ (\AA)& 4.18 & 3.59 & 4.12 & 3.86 \\
 \hline
 $\bf{k_{bind}}$ (eV/\AA$^2$) & -0.38 & -0.86 & -0.31 & -0.45 \\
 \hline
\end{tabular}
\caption{Results of fitting DFT total energy data to Rose's universal binding potential \cite{rose1981universal}. 
}
\end{table}

In agreement with the results in Fig. 4, in the case of Au, the breaking force is twice as high when relativistic effects are included. For silver, however, it is only about $25 \%$ larger. Moreover, since the fitted inter-electrode binding constant $k_{\mathrm{bind}}$ of relativistic Au is at least twice as large as in all the other cases, taking into account the intra-electrode elasticity arguments presented earlier, it is to be expected that relativistic Au would exhibit a larger jump to contact. This is precisely what we observe in our experimental data.

In summary, we have reported a direct measurement of the strong relativistic effect in the formation of single-atom gold contacts. This phenomenon was revealed by the introduction of a new statistical treatment of the experimental data, and can be fully understood from a comparison of the experiments with DFT calculations in which scalar relativistic corrections are included or not.

\section{ACKNOWLEDGMENTS}
This work has been funded by the Spanish MEC through grants FIS2013-47328 and MAT2016-78625. 
C.S. gratefully acknowledges financial support from SEPE Servicio P\'ublico de Empleo Estatal. W.D. acknowledges funding from the National Research Foundation of South Africa through the Innovation Doctoral scholarship programme, Grant UID 102574. W.D. also thanks Prof. A. E. Botha for sharing the python scripts used in the polynomial interpolations and E. Artacho, J. Fern\'andez-Rossier and J.J. Palacios for fruitful discussions. The DFT calculations in this paper were performed on the high-performance computing (HPC) facility at UNISA.

\bibliography{j2c_relativistic__1_.bib,RelativisticJ2C.bib}

\begin{thebibliography}{32}%
\makeatletter
\providecommand \@ifxundefined [1]{%
 \@ifx{#1\undefined}
}%
\providecommand \@ifnum [1]{%
 \ifnum #1\expandafter \@firstoftwo
 \else \expandafter \@secondoftwo
 \fi
}%
\providecommand \@ifx [1]{%
 \ifx #1\expandafter \@firstoftwo
 \else \expandafter \@secondoftwo
 \fi
}%
\providecommand \natexlab [1]{#1}%
\providecommand \enquote  [1]{``#1''}%
\providecommand \bibnamefont  [1]{#1}%
\providecommand \bibfnamefont [1]{#1}%
\providecommand \citenamefont [1]{#1}%
\providecommand \href@noop [0]{\@secondoftwo}%
\providecommand \href [0]{\begingroup \@sanitize@url \@href}%
\providecommand \@href[1]{\@@startlink{#1}\@@href}%
\providecommand \@@href[1]{\endgroup#1\@@endlink}%
\providecommand \@sanitize@url [0]{\catcode `\\12\catcode `\$12\catcode
  `\&12\catcode `\#12\catcode `\^12\catcode `\_12\catcode `\%12\relax}%
\providecommand \@@startlink[1]{}%
\providecommand \@@endlink[0]{}%
\providecommand \url  [0]{\begingroup\@sanitize@url \@url }%
\providecommand \@url [1]{\endgroup\@href {#1}{\urlprefix }}%
\providecommand \urlprefix  [0]{URL }%
\providecommand \Eprint [0]{\href }%
\providecommand \doibase [0]{http://dx.doi.org/}%
\providecommand \selectlanguage [0]{\@gobble}%
\providecommand \bibinfo  [0]{\@secondoftwo}%
\providecommand \bibfield  [0]{\@secondoftwo}%
\providecommand \translation [1]{[#1]}%
\providecommand \BibitemOpen [0]{}%
\providecommand \bibitemStop [0]{}%
\providecommand \bibitemNoStop [0]{.\EOS\space}%
\providecommand \EOS [0]{\spacefactor3000\relax}%
\providecommand \BibitemShut  [1]{\csname bibitem#1\endcsname}%
\let\auto@bib@innerbib\@empty
\bibitem [{\citenamefont {Pyykko}(1988)}]{pyykko1988relativistic}%
  \BibitemOpen
  \bibfield  {author} {\bibinfo {author} {\bibfnamefont {P.}~\bibnamefont
  {Pyykko}},\ }\href@noop {} {\bibfield  {journal} {\bibinfo  {journal}
  {Chemical Reviews}\ }\textbf {\bibinfo {volume} {88}},\ \bibinfo {pages}
  {563} (\bibinfo {year} {1988})}\BibitemShut {NoStop}%
\bibitem [{\citenamefont {Calvo}\ \emph {et~al.}(2013)\citenamefont {Calvo},
  \citenamefont {Pahl}, \citenamefont {Wormit},\ and\ \citenamefont
  {Schwerdtfeger}}]{Calvo_2013}%
  \BibitemOpen
  \bibfield  {author} {\bibinfo {author} {\bibfnamefont {F.}~\bibnamefont
  {Calvo}}, \bibinfo {author} {\bibfnamefont {E.}~\bibnamefont {Pahl}},
  \bibinfo {author} {\bibfnamefont {M.}~\bibnamefont {Wormit}}, \ and\ \bibinfo
  {author} {\bibfnamefont {P.}~\bibnamefont {Schwerdtfeger}},\ }\href {\doibase
  10.1002/anie.201302742} {\bibfield  {journal} {\bibinfo  {journal}
  {Angewandte Chemie International Edition}\ }\textbf {\bibinfo {volume}
  {52}},\ \bibinfo {pages} {7583} (\bibinfo {year} {2013})}\BibitemShut
  {NoStop}%
\bibitem [{\citenamefont {Gorin}\ and\ \citenamefont
  {Toste}(2007)}]{Gorin_2007}%
  \BibitemOpen
  \bibfield  {author} {\bibinfo {author} {\bibfnamefont {D.~J.}\ \bibnamefont
  {Gorin}}\ and\ \bibinfo {author} {\bibfnamefont {F.~D.}\ \bibnamefont
  {Toste}},\ }\href {\doibase 10.1038/nature05592} {\bibfield  {journal}
  {\bibinfo  {journal} {Nature}\ }\textbf {\bibinfo {volume} {446}},\ \bibinfo
  {pages} {395} (\bibinfo {year} {2007})}\BibitemShut {NoStop}%
\bibitem [{\citenamefont {Ho}\ and\ \citenamefont
  {Bohnen}(1987)}]{ho1987stability}%
  \BibitemOpen
  \bibfield  {author} {\bibinfo {author} {\bibfnamefont {K.~M.}\ \bibnamefont
  {Ho}}\ and\ \bibinfo {author} {\bibfnamefont {K.~P.}\ \bibnamefont
  {Bohnen}},\ }\href {\doibase 10.1103/PhysRevLett.59.1833} {\bibfield
  {journal} {\bibinfo  {journal} {Phys. Rev. Lett.}\ }\textbf {\bibinfo
  {volume} {59}},\ \bibinfo {pages} {1833} (\bibinfo {year}
  {1987})}\BibitemShut {NoStop}%
\bibitem [{\citenamefont {Takeuchi}\ \emph {et~al.}(1989)\citenamefont
  {Takeuchi}, \citenamefont {Chan},\ and\ \citenamefont
  {Ho}}]{takeuchi1989theoretical}%
  \BibitemOpen
  \bibfield  {author} {\bibinfo {author} {\bibfnamefont {N.}~\bibnamefont
  {Takeuchi}}, \bibinfo {author} {\bibfnamefont {C.~T.}\ \bibnamefont {Chan}},
  \ and\ \bibinfo {author} {\bibfnamefont {K.~M.}\ \bibnamefont {Ho}},\ }\href
  {\doibase 10.1103/PhysRevLett.63.1273} {\bibfield  {journal} {\bibinfo
  {journal} {Phys. Rev. Lett.}\ }\textbf {\bibinfo {volume} {63}},\ \bibinfo
  {pages} {1273} (\bibinfo {year} {1989})}\BibitemShut {NoStop}%
\bibitem [{\citenamefont {Filippetti}\ and\ \citenamefont
  {Fiorentini}(1997)}]{filippetti1997reconstructions}%
  \BibitemOpen
  \bibfield  {author} {\bibinfo {author} {\bibfnamefont {A.}~\bibnamefont
  {Filippetti}}\ and\ \bibinfo {author} {\bibfnamefont {V.}~\bibnamefont
  {Fiorentini}},\ }\href@noop {} {\bibfield  {journal} {\bibinfo  {journal}
  {Surface science}\ }\textbf {\bibinfo {volume} {377}},\ \bibinfo {pages}
  {112} (\bibinfo {year} {1997})}\BibitemShut {NoStop}%
\bibitem [{\citenamefont {Smit}\ \emph {et~al.}(2001)\citenamefont {Smit},
  \citenamefont {Untiedt}, \citenamefont {Yanson},\ and\ \citenamefont {van
  Ruitenbeek}}]{Smit_2001}%
  \BibitemOpen
  \bibfield  {author} {\bibinfo {author} {\bibfnamefont {R.~H.~M.}\
  \bibnamefont {Smit}}, \bibinfo {author} {\bibfnamefont {C.}~\bibnamefont
  {Untiedt}}, \bibinfo {author} {\bibfnamefont {A.~I.}\ \bibnamefont {Yanson}},
  \ and\ \bibinfo {author} {\bibfnamefont {J.~M.}\ \bibnamefont {van
  Ruitenbeek}},\ }\href {\doibase 10.1103/PhysRevLett.87.266102} {\bibfield
  {journal} {\bibinfo  {journal} {Phys. Rev. Lett.}\ }\textbf {\bibinfo
  {volume} {87}},\ \bibinfo {pages} {266102} (\bibinfo {year}
  {2001})}\BibitemShut {NoStop}%
\bibitem [{\citenamefont {H{\"a}kkinen}\ \emph {et~al.}(2000)\citenamefont
  {H{\"a}kkinen}, \citenamefont {Barnett}, \citenamefont {Scherbakov},\ and\
  \citenamefont {Landman}}]{hakkinen2000nanowire}%
  \BibitemOpen
  \bibfield  {author} {\bibinfo {author} {\bibfnamefont {H.}~\bibnamefont
  {H{\"a}kkinen}}, \bibinfo {author} {\bibfnamefont {R.~N.}\ \bibnamefont
  {Barnett}}, \bibinfo {author} {\bibfnamefont {A.~G.}\ \bibnamefont
  {Scherbakov}}, \ and\ \bibinfo {author} {\bibfnamefont {U.}~\bibnamefont
  {Landman}},\ }\href@noop {} {\bibfield  {journal} {\bibinfo  {journal} {The
  Journal of Physical Chemistry B}\ }\textbf {\bibinfo {volume} {104}},\
  \bibinfo {pages} {9063} (\bibinfo {year} {2000})}\BibitemShut {NoStop}%
\bibitem [{\citenamefont {Bahn}\ and\ \citenamefont
  {Jacobsen}(2001)}]{Bahn_2001}%
  \BibitemOpen
  \bibfield  {author} {\bibinfo {author} {\bibfnamefont {S.~R.}\ \bibnamefont
  {Bahn}}\ and\ \bibinfo {author} {\bibfnamefont {K.~W.}\ \bibnamefont
  {Jacobsen}},\ }\href {\doibase 10.1103/PhysRevLett.87.266101} {\bibfield
  {journal} {\bibinfo  {journal} {Phys. Rev. Lett.}\ }\textbf {\bibinfo
  {volume} {87}},\ \bibinfo {pages} {266101} (\bibinfo {year}
  {2001})}\BibitemShut {NoStop}%
\bibitem [{\citenamefont {Hasmy}\ \emph {et~al.}(2008)\citenamefont {Hasmy},
  \citenamefont {Rinc\'on}, \citenamefont {Hern\'andez}, \citenamefont
  {Mujica}, \citenamefont {M\'arquez},\ and\ \citenamefont
  {Gonz\'alez}}]{Hasmy_2008}%
  \BibitemOpen
  \bibfield  {author} {\bibinfo {author} {\bibfnamefont {A.}~\bibnamefont
  {Hasmy}}, \bibinfo {author} {\bibfnamefont {L.}~\bibnamefont {Rinc\'on}},
  \bibinfo {author} {\bibfnamefont {R.}~\bibnamefont {Hern\'andez}}, \bibinfo
  {author} {\bibfnamefont {V.}~\bibnamefont {Mujica}}, \bibinfo {author}
  {\bibfnamefont {M.}~\bibnamefont {M\'arquez}}, \ and\ \bibinfo {author}
  {\bibfnamefont {C.}~\bibnamefont {Gonz\'alez}},\ }\href {\doibase
  10.1103/PhysRevB.78.115409} {\bibfield  {journal} {\bibinfo  {journal} {Phys.
  Rev. B}\ }\textbf {\bibinfo {volume} {78}},\ \bibinfo {pages} {115409}
  (\bibinfo {year} {2008})}\BibitemShut {NoStop}%
\bibitem [{\citenamefont {Thiess}\ \emph {et~al.}(2008)\citenamefont {Thiess},
  \citenamefont {Mokrousov}, \citenamefont {Blugel},\ and\ \citenamefont
  {Heinze}}]{thiess2008theory}%
  \BibitemOpen
  \bibfield  {author} {\bibinfo {author} {\bibfnamefont {A.}~\bibnamefont
  {Thiess}}, \bibinfo {author} {\bibfnamefont {Y.}~\bibnamefont {Mokrousov}},
  \bibinfo {author} {\bibfnamefont {S.}~\bibnamefont {Blugel}}, \ and\ \bibinfo
  {author} {\bibfnamefont {S.}~\bibnamefont {Heinze}},\ }\href@noop {}
  {\bibfield  {journal} {\bibinfo  {journal} {Nano letters}\ }\textbf {\bibinfo
  {volume} {8}},\ \bibinfo {pages} {2144} (\bibinfo {year} {2008})}\BibitemShut
  {NoStop}%
\bibitem [{\citenamefont {Agra{\" \i}t}\ \emph {et~al.}(2003)\citenamefont
  {Agra{\" \i}t}, \citenamefont {Yeyati},\ and\ \citenamefont
  {Van~Ruitenbeek}}]{agrait2003quantum}%
  \BibitemOpen
  \bibfield  {author} {\bibinfo {author} {\bibfnamefont {N.}~\bibnamefont
  {Agra{\" \i}t}}, \bibinfo {author} {\bibfnamefont {A.~L.}\ \bibnamefont
  {Yeyati}}, \ and\ \bibinfo {author} {\bibfnamefont {J.~M.}\ \bibnamefont
  {Van~Ruitenbeek}},\ }\href@noop {} {\bibfield  {journal} {\bibinfo  {journal}
  {Physics Reports}\ }\textbf {\bibinfo {volume} {377}},\ \bibinfo {pages} {81}
  (\bibinfo {year} {2003})}\BibitemShut {NoStop}%
\bibitem [{\citenamefont {Kr{\"o}ger}\ \emph {et~al.}(2009)\citenamefont
  {Kr{\"o}ger}, \citenamefont {N{\'e}el}, \citenamefont {Sperl}, \citenamefont
  {Wang},\ and\ \citenamefont {Berndt}}]{Kr_ger_2009}%
  \BibitemOpen
  \bibfield  {author} {\bibinfo {author} {\bibfnamefont {J.}~\bibnamefont
  {Kr{\"o}ger}}, \bibinfo {author} {\bibfnamefont {N.}~\bibnamefont
  {N{\'e}el}}, \bibinfo {author} {\bibfnamefont {A.}~\bibnamefont {Sperl}},
  \bibinfo {author} {\bibfnamefont {Y.~F.}\ \bibnamefont {Wang}}, \ and\
  \bibinfo {author} {\bibfnamefont {R.}~\bibnamefont {Berndt}},\ }\href
  {\doibase 10.1088/1367-2630/11/12/125006} {\bibfield  {journal} {\bibinfo
  {journal} {New Journal of Physics}\ }\textbf {\bibinfo {volume} {11}},\
  \bibinfo {pages} {125006} (\bibinfo {year} {2009})}\BibitemShut {NoStop}%
\bibitem [{\citenamefont {Untiedt}\ \emph {et~al.}(2007)\citenamefont
  {Untiedt}, \citenamefont {Caturla}, \citenamefont {Calvo}, \citenamefont
  {Palacios}, \citenamefont {Segers},\ and\ \citenamefont {van
  Ruitenbeek}}]{Untiedt_2007}%
  \BibitemOpen
  \bibfield  {author} {\bibinfo {author} {\bibfnamefont {C.}~\bibnamefont
  {Untiedt}}, \bibinfo {author} {\bibfnamefont {M.~J.}\ \bibnamefont
  {Caturla}}, \bibinfo {author} {\bibfnamefont {M.~R.}\ \bibnamefont {Calvo}},
  \bibinfo {author} {\bibfnamefont {J.~J.}\ \bibnamefont {Palacios}}, \bibinfo
  {author} {\bibfnamefont {R.~C.}\ \bibnamefont {Segers}}, \ and\ \bibinfo
  {author} {\bibfnamefont {J.~M.}\ \bibnamefont {van Ruitenbeek}},\ }\href
  {\doibase 10.1103/PhysRevLett.98.206801} {\bibfield  {journal} {\bibinfo
  {journal} {Phys. Rev. Lett.}\ }\textbf {\bibinfo {volume} {98}},\ \bibinfo
  {pages} {206801} (\bibinfo {year} {2007})}\BibitemShut {NoStop}%
\bibitem [{\citenamefont {Sabater}\ \emph {et~al.}(2013)\citenamefont
  {Sabater}, \citenamefont {Caturla}, \citenamefont {Palacios},\ and\
  \citenamefont {Untiedt}}]{sabater2013understanding}%
  \BibitemOpen
  \bibfield  {author} {\bibinfo {author} {\bibfnamefont {C.}~\bibnamefont
  {Sabater}}, \bibinfo {author} {\bibfnamefont {M.~J.}\ \bibnamefont
  {Caturla}}, \bibinfo {author} {\bibfnamefont {J.~J.}\ \bibnamefont
  {Palacios}}, \ and\ \bibinfo {author} {\bibfnamefont {C.}~\bibnamefont
  {Untiedt}},\ }\href@noop {} {\bibfield  {journal} {\bibinfo  {journal}
  {Nanoscale research letters}\ }\textbf {\bibinfo {volume} {8}},\ \bibinfo
  {pages} {257} (\bibinfo {year} {2013})}\BibitemShut {NoStop}%
\bibitem [{\citenamefont {Limot}\ \emph {et~al.}(2005)\citenamefont {Limot},
  \citenamefont {Kr\"oger}, \citenamefont {Berndt}, \citenamefont
  {Garcia-Lekue},\ and\ \citenamefont {Hofer}}]{Limot_2005}%
  \BibitemOpen
  \bibfield  {author} {\bibinfo {author} {\bibfnamefont {L.}~\bibnamefont
  {Limot}}, \bibinfo {author} {\bibfnamefont {J.}~\bibnamefont {Kr\"oger}},
  \bibinfo {author} {\bibfnamefont {R.}~\bibnamefont {Berndt}}, \bibinfo
  {author} {\bibfnamefont {A.}~\bibnamefont {Garcia-Lekue}}, \ and\ \bibinfo
  {author} {\bibfnamefont {W.~A.}\ \bibnamefont {Hofer}},\ }\href {\doibase
  10.1103/PhysRevLett.94.126102} {\bibfield  {journal} {\bibinfo  {journal}
  {Phys. Rev. Lett.}\ }\textbf {\bibinfo {volume} {94}},\ \bibinfo {pages}
  {126102} (\bibinfo {year} {2005})}\BibitemShut {NoStop}%
\bibitem [{\citenamefont {Trouwborst}\ \emph {et~al.}(2008)\citenamefont
  {Trouwborst}, \citenamefont {Huisman}, \citenamefont {Bakker}, \citenamefont
  {van~der Molen},\ and\ \citenamefont {van Wees}}]{Trouwborst_2008}%
  \BibitemOpen
  \bibfield  {author} {\bibinfo {author} {\bibfnamefont {M.~L.}\ \bibnamefont
  {Trouwborst}}, \bibinfo {author} {\bibfnamefont {E.~H.}\ \bibnamefont
  {Huisman}}, \bibinfo {author} {\bibfnamefont {F.~L.}\ \bibnamefont {Bakker}},
  \bibinfo {author} {\bibfnamefont {S.~J.}\ \bibnamefont {van~der Molen}}, \
  and\ \bibinfo {author} {\bibfnamefont {B.~J.}\ \bibnamefont {van Wees}},\
  }\href {\doibase 10.1103/PhysRevLett.100.175502} {\bibfield  {journal}
  {\bibinfo  {journal} {Phys. Rev. Lett.}\ }\textbf {\bibinfo {volume} {100}},\
  \bibinfo {pages} {175502} (\bibinfo {year} {2008})}\BibitemShut {NoStop}%
\bibitem [{\citenamefont {Fern\'andez}\ \emph {et~al.}(2016)\citenamefont
  {Fern\'andez}, \citenamefont {Sabater}, \citenamefont {Dednam}, \citenamefont
  {Palacios}, \citenamefont {Calvo}, \citenamefont {Untiedt},\ and\
  \citenamefont {Caturla}}]{Fern_ndez_2016}%
  \BibitemOpen
  \bibfield  {author} {\bibinfo {author} {\bibfnamefont {M.~A.}\ \bibnamefont
  {Fern\'andez}}, \bibinfo {author} {\bibfnamefont {C.}~\bibnamefont
  {Sabater}}, \bibinfo {author} {\bibfnamefont {W.}~\bibnamefont {Dednam}},
  \bibinfo {author} {\bibfnamefont {J.~J.}\ \bibnamefont {Palacios}}, \bibinfo
  {author} {\bibfnamefont {M.~R.}\ \bibnamefont {Calvo}}, \bibinfo {author}
  {\bibfnamefont {C.}~\bibnamefont {Untiedt}}, \ and\ \bibinfo {author}
  {\bibfnamefont {M.~J.}\ \bibnamefont {Caturla}},\ }\href {\doibase
  10.1103/PhysRevB.93.085437} {\bibfield  {journal} {\bibinfo  {journal} {Phys.
  Rev. B}\ }\textbf {\bibinfo {volume} {93}},\ \bibinfo {pages} {085437}
  (\bibinfo {year} {2016})}\BibitemShut {NoStop}%
\bibitem [{\citenamefont {Scheer}\ \emph {et~al.}(1998)\citenamefont {Scheer},
  \citenamefont {Agra{\"\i}t}, \citenamefont {Cuevas}, \citenamefont {Yeyati},
  \citenamefont {Ludoph}, \citenamefont {Mart{\'\i}n-Rodero}, \citenamefont
  {Bollinger}, \citenamefont {van Ruitenbeek},\ and\ \citenamefont
  {Urbina}}]{scheer1998signature}%
  \BibitemOpen
  \bibfield  {author} {\bibinfo {author} {\bibfnamefont {E.}~\bibnamefont
  {Scheer}}, \bibinfo {author} {\bibfnamefont {N.}~\bibnamefont {Agra{\"\i}t}},
  \bibinfo {author} {\bibfnamefont {J.~C.}\ \bibnamefont {Cuevas}}, \bibinfo
  {author} {\bibfnamefont {A.~L.}\ \bibnamefont {Yeyati}}, \bibinfo {author}
  {\bibfnamefont {B.}~\bibnamefont {Ludoph}}, \bibinfo {author} {\bibfnamefont
  {A.}~\bibnamefont {Mart{\'\i}n-Rodero}}, \bibinfo {author} {\bibfnamefont
  {G.~R.}\ \bibnamefont {Bollinger}}, \bibinfo {author} {\bibfnamefont {J.~M.}\
  \bibnamefont {van Ruitenbeek}}, \ and\ \bibinfo {author} {\bibfnamefont
  {C.}~\bibnamefont {Urbina}},\ }\href@noop {} {\bibfield  {journal} {\bibinfo
  {journal} {Nature}\ }\textbf {\bibinfo {volume} {394}},\ \bibinfo {pages}
  {154} (\bibinfo {year} {1998})}\BibitemShut {NoStop}%
\bibitem [{Note1()}]{Note1}%
  \BibitemOpen
  \bibinfo {note} {Log denotes here the common logarithm (base 10)}\BibitemShut
  {NoStop}%
\bibitem [{PRB()}]{PRB}%
  \BibitemOpen
  \href@noop {} {\ }\bibinfo {note} {Associated article submitted to
  PRB}\BibitemShut {NoStop}%
\bibitem [{\citenamefont {Hybertsen}\ and\ \citenamefont
  {Venkataraman}(2016)}]{Hybertsen_2016}%
  \BibitemOpen
  \bibfield  {author} {\bibinfo {author} {\bibfnamefont {M.~S.}\ \bibnamefont
  {Hybertsen}}\ and\ \bibinfo {author} {\bibfnamefont {L.}~\bibnamefont
  {Venkataraman}},\ }\href {\doibase 10.1021/acs.accounts.6b00004} {\bibfield
  {journal} {\bibinfo  {journal} {Accounts of Chemical Research}\ }\textbf
  {\bibinfo {volume} {49}},\ \bibinfo {pages} {452} (\bibinfo {year}
  {2016})}\BibitemShut {NoStop}%
\bibitem [{\citenamefont {Clark}\ \emph {et~al.}(2005)\citenamefont {Clark},
  \citenamefont {Segall}, \citenamefont {Pickard}, \citenamefont {Hasnip},
  \citenamefont {Probert}, \citenamefont {Refson},\ and\ \citenamefont
  {Payne}}]{clark2005first}%
  \BibitemOpen
  \bibfield  {author} {\bibinfo {author} {\bibfnamefont {S.~J.}\ \bibnamefont
  {Clark}}, \bibinfo {author} {\bibfnamefont {M.~D.}\ \bibnamefont {Segall}},
  \bibinfo {author} {\bibfnamefont {C.~J.}\ \bibnamefont {Pickard}}, \bibinfo
  {author} {\bibfnamefont {P.~J.}\ \bibnamefont {Hasnip}}, \bibinfo {author}
  {\bibfnamefont {M.~I.}\ \bibnamefont {Probert}}, \bibinfo {author}
  {\bibfnamefont {K.}~\bibnamefont {Refson}}, \ and\ \bibinfo {author}
  {\bibfnamefont {M.~C.}\ \bibnamefont {Payne}},\ }\href@noop {} {\bibfield
  {journal} {\bibinfo  {journal} {Zeitschrift f{\"u}r
  Kristallographie-Crystalline Materials}\ }\textbf {\bibinfo {volume} {220}},\
  \bibinfo {pages} {567} (\bibinfo {year} {2005})}\BibitemShut {NoStop}%
\bibitem [{\citenamefont {Vanderbilt}(1990)}]{Vanderbilt_1990}%
  \BibitemOpen
  \bibfield  {author} {\bibinfo {author} {\bibfnamefont {D.}~\bibnamefont
  {Vanderbilt}},\ }\href {\doibase 10.1103/PhysRevB.41.7892} {\bibfield
  {journal} {\bibinfo  {journal} {Phys. Rev. B}\ }\textbf {\bibinfo {volume}
  {41}},\ \bibinfo {pages} {7892} (\bibinfo {year} {1990})}\BibitemShut
  {NoStop}%
\bibitem [{\citenamefont {Lejaeghere}\ \emph {et~al.}(2013)\citenamefont
  {Lejaeghere}, \citenamefont {Van~Speybroeck}, \citenamefont {Van~Oost},\ and\
  \citenamefont {Cottenier}}]{Lejaeghere_2013}%
  \BibitemOpen
  \bibfield  {author} {\bibinfo {author} {\bibfnamefont {K.}~\bibnamefont
  {Lejaeghere}}, \bibinfo {author} {\bibfnamefont {V.}~\bibnamefont
  {Van~Speybroeck}}, \bibinfo {author} {\bibfnamefont {G.}~\bibnamefont
  {Van~Oost}}, \ and\ \bibinfo {author} {\bibfnamefont {S.}~\bibnamefont
  {Cottenier}},\ }\href {\doibase 10.1080/10408436.2013.772503} {\bibfield
  {journal} {\bibinfo  {journal} {Critical Reviews in Solid State and Materials
  Sciences}\ }\textbf {\bibinfo {volume} {39}},\ \bibinfo {pages} {1} (\bibinfo
  {year} {2013})}\BibitemShut {NoStop}%
\bibitem [{\citenamefont {Koelling}\ and\ \citenamefont
  {Harmon}(1977)}]{Koelling_1977}%
  \BibitemOpen
  \bibfield  {author} {\bibinfo {author} {\bibfnamefont {D.~D.}\ \bibnamefont
  {Koelling}}\ and\ \bibinfo {author} {\bibfnamefont {B.~N.}\ \bibnamefont
  {Harmon}},\ }\href {\doibase 10.1088/0022-3719/10/16/019} {\bibfield
  {journal} {\bibinfo  {journal} {J. Phys. C: Solid St. Phys.}\ }\textbf
  {\bibinfo {volume} {10}},\ \bibinfo {pages} {3107} (\bibinfo {year}
  {1977})}\BibitemShut {NoStop}%
\bibitem [{\citenamefont {Strigl}\ \emph {et~al.}(2015)\citenamefont {Strigl}
  \emph {et~al.}}]{Strigl_2015}%
  \BibitemOpen
  \bibfield  {author} {\bibinfo {author} {\bibfnamefont {F.}~\bibnamefont
  {Strigl}} \emph {et~al.},\ }\href {\doibase 10.1038/ncomms7172} {\bibfield
  {journal} {\bibinfo  {journal} {Nat. Commun.}\ }\textbf {\bibinfo {volume}
  {6}},\ \bibinfo {pages} {6172} (\bibinfo {year} {2015})}\BibitemShut
  {NoStop}%
\bibitem [{\citenamefont {Perdew}\ \emph {et~al.}(1996)\citenamefont {Perdew},
  \citenamefont {Burke},\ and\ \citenamefont
  {Ernzerhof}}]{perdew1996generalized}%
  \BibitemOpen
  \bibfield  {author} {\bibinfo {author} {\bibfnamefont {J.~P.}\ \bibnamefont
  {Perdew}}, \bibinfo {author} {\bibfnamefont {K.}~\bibnamefont {Burke}}, \
  and\ \bibinfo {author} {\bibfnamefont {M.}~\bibnamefont {Ernzerhof}},\
  }\href@noop {} {\bibfield  {journal} {\bibinfo  {journal} {Physical review
  letters}\ }\textbf {\bibinfo {volume} {77}},\ \bibinfo {pages} {3865}
  (\bibinfo {year} {1996})}\BibitemShut {NoStop}%
\bibitem [{\citenamefont {Hybertsen}(2017)}]{Hybertsen_2017}%
  \BibitemOpen
  \bibfield  {author} {\bibinfo {author} {\bibfnamefont {M.~S.}\ \bibnamefont
  {Hybertsen}},\ }\href {\doibase 10.1063/1.4975769} {\bibfield  {journal}
  {\bibinfo  {journal} {The Journal of Chemical Physics}\ }\textbf {\bibinfo
  {volume} {146}},\ \bibinfo {pages} {092323} (\bibinfo {year}
  {2017})}\BibitemShut {NoStop}%
\bibitem [{\citenamefont {Barzilai}\ and\ \citenamefont
  {Borwein}(1988)}]{barzilai1988two}%
  \BibitemOpen
  \bibfield  {author} {\bibinfo {author} {\bibfnamefont {J.}~\bibnamefont
  {Barzilai}}\ and\ \bibinfo {author} {\bibfnamefont {J.~M.}\ \bibnamefont
  {Borwein}},\ }\href@noop {} {\bibfield  {journal} {\bibinfo  {journal} {IMA
  journal of numerical analysis}\ }\textbf {\bibinfo {volume} {8}},\ \bibinfo
  {pages} {141} (\bibinfo {year} {1988})}\BibitemShut {NoStop}%
\bibitem [{\citenamefont {Hofer}\ \emph {et~al.}(2004)\citenamefont {Hofer},
  \citenamefont {Garcia-Lekue},\ and\ \citenamefont {Brune}}]{Hofer_2004}%
  \BibitemOpen
  \bibfield  {author} {\bibinfo {author} {\bibfnamefont {W.}~\bibnamefont
  {Hofer}}, \bibinfo {author} {\bibfnamefont {A.}~\bibnamefont {Garcia-Lekue}},
  \ and\ \bibinfo {author} {\bibfnamefont {H.}~\bibnamefont {Brune}},\ }\href
  {\doibase 10.1016/j.cplett.2004.08.110} {\bibfield  {journal} {\bibinfo
  {journal} {Chemical Physics Letters}\ }\textbf {\bibinfo {volume} {397}},\
  \bibinfo {pages} {354} (\bibinfo {year} {2004})}\BibitemShut {NoStop}%
\bibitem [{\citenamefont {Rose}\ \emph {et~al.}(1981)\citenamefont {Rose},
  \citenamefont {Ferrante},\ and\ \citenamefont {Smith}}]{rose1981universal}%
  \BibitemOpen
  \bibfield  {author} {\bibinfo {author} {\bibfnamefont {J.~H.}\ \bibnamefont
  {Rose}}, \bibinfo {author} {\bibfnamefont {J.}~\bibnamefont {Ferrante}}, \
  and\ \bibinfo {author} {\bibfnamefont {J.~R.}\ \bibnamefont {Smith}},\
  }\href@noop {} {\bibfield  {journal} {\bibinfo  {journal} {Physical Review
  Letters}\ }\textbf {\bibinfo {volume} {47}},\ \bibinfo {pages} {675}
  (\bibinfo {year} {1981})}\BibitemShut {NoStop}%
\end{thebibliography}%








\end{document}